\documentclass[trackchanges, twocolumn, shortbib, resetfootnote]{aastex701}
\usepackage{amsmath, amssymb}
\usepackage{multirow}
\usepackage{graphicx,color,float,tabularx,siunitx}
\definecolor{colorLink}{rgb}{0,0,180} 
\usepackage{hyperref}
\hypersetup{
   colorlinks = true,
   citecolor  = colorLink,
   urlcolor   = colorLink,
   linkcolor  = colorLink,
}

\begin{document}

\title{Reconstructing the Auger UHECR Dipole: A Hybrid Analysis of 4LAC AGNs and Nearby Starburst Galaxies}

\author{Swaraj Pratim Sarmah}
\affiliation{Department of Physics, Indian Institute of Technology, Guwahati, 781039, Assam, India}
\email{swarajpratimsarmah16@gmail.com}

\author{Umananda Dev Goswami}
\affiliation{Department of Physics, Dibrugarh University, Dibrugarh, 786004, Assam, India}
\email{umananda@dibru.ac.in}

\author{Sovan Chakraborty}
\affiliation{Department of Physics, Indian Institute of Technology, Guwahati, 781039, Assam, India}
\email{sovan@iitg.ac.in}

\begin{abstract}
The large-scale dipole anisotropy observed by the Pierre Auger Observatory 
above $8$~EeV provides important clues about the origin of ultra-high-energy 
cosmic rays (UHECRs). In this work, we investigate whether gamma-ray active 
galactic nuclei (AGNs) from the Fermi-LAT Fourth AGN Catalog (4LAC) can 
explain the observed dipole and examine the contribution of nearby starburst 
galaxies (SBGs). Incorporating source fluxes, redshift distributions, GZK 
attenuation, and a mixed-composition framework, we demonstrate that 
single-source populations of AGNs and  SBGs alone fail to reproduce the 
observed Auger dipole. To resolve this, we construct a hybrid AGN+SBG model. 
The best-fit solution for the $8$--$16$~EeV interval requires a $23$\% AGN 
contribution and $45$\% SBG contribution, which gives a dipole amplitude of $5.03$\% while maintaining the 
dipole direction of ($104.1^\circ$, $-25.1^\circ$) with angular separation of 
$7.10^\circ$. In the other energy intervals also, we find that the nearby SBGs increasingly dominate the anisotropic component while 
AGNs provide a subdominant contribution alongside a quasi-isotropic background.
\end{abstract}


\section{Introduction} \label{sec:intro}

The arrival directions of studying ultra-high-energy cosmic ray (UHECR) provides crucial information into their astrophysical origin. While searches above $\sim 40$~EeV show only tentative hints of intermediate-scale ($\sim 10^\circ$--$20^\circ$) anisotropy \citep{PierreAuger:2014yba, PierreAuger:2018qvk}, the Pierre Auger Collaboration confirmed a large-scale dipole above $8$~EeV \citep{Auger2017a}. This $\sim 6.5\%$ dipole points roughly $125^\circ$ away from the Galactic Center, establishing an extragalactic origin \citep{PierreAuger:2018zqu}. The lack of small-scale clustering alongside this clear dipole points to strong magnetic deflections, consistent with cosmic rays becoming heavier at higher energies \citep{auger_prd9012, auger_prd90, Auger2017}. Additionally, deflections in the Galactic magnetic field modify both the amplitude and observed direction of the dipole \citep{PierreAuger:2018zqu}.

UHECR fluxes and compositions are expected to be significantly modified by propagation 
effects, notably the Greisen--Zatsepin--Kuzmin (GZK) suppression 
caused 
by cosmic microwave background (CMB) photons \citep{KGreisen1966, GZ1966, 
Dermer:2008cy}, with primary compositions remaining uncertain at the highest 
energies \citep{HiRes:2009fiy}. Proposed sources are highly diverse, ranging 
from ultraheavy nuclei \citep{Zhang:2024sjp}, binary neutron star mergers 
\citep{Farrar:2024zsm}, and unresolved transients \citep{Unger:2023hnu} to 
decaying superheavy dark matter \citep{Murase:2025uwv}. There are also 
proposed reasonable effects of modified gravity frameworks on their fluxes and
composition \citep{Sarmah:2025uzk, Sarmah:2024kek}. UHECRs also serve 
as sensitive probes for new physics, such as Lorentz invariance violation 
\citep{auger2022, Lang:2024jmc}.

The UHECR energy generation rate above $10^{18}$~eV is $\sim 10^{45}\text{ erg Mpc}^{-3}\text{ yr}^{-1}$ \citep{Unger:2015laa}, matching the gamma-ray output of AGNs and SBGs \citep{Dermer:2010iz}. Due to their low space density and strong GZK losses, a few nearby sources should dominate the flux, naturally driving intermediate-scale anisotropy \citep{PierreAuger:2018qvk}. Both classes are physically well-motivated: AGNs satisfy the Hillas criterion for shock acceleration \citep{Hillas:1984ijl}, while SBGs have elevated rates of extreme stellar explosions such as GRBs, hypernovae, and magnetars \citep{Biermann:2016xzl, Perley:2016bke}.

The observed UHECR energy generation rate above $10^{18}$~eV is $\sim 10^{45}\text{ erg Mpc}^{-3}\text{ yr}^{-1}$ \citep{Unger:2015laa}, matching the integrated gamma-ray output of both active galactic nuclei (AGNs) and starburst galaxies (SBGs) \citep{Dermer:2010iz}. Both source classes have solid physical motivation: AGNs satisfy the Hillas criterion for shock acceleration in relativistic jets \citep{Hillas:1984ijl}, whereas SBGs host elevated rates of extreme stellar explosions, including gamma-ray bursts, hypernovae, and magnetars \citep{Biermann:2016xzl, Perley:2016bke}. Due to their low space density and strong propagation losses via GZK attenuation, a small number of nearby objects are expected to dominate the local flux, naturally producing anisotropy \citep{PierreAuger:2018qvk}.
Consequently, nearby radio and starburst galaxies are currently favoured as dominant contributors to the highest-energy flux and anisotropy of UHECRs \citep{Seo:2025oiw, Eichmann:2022ias, Wang:2024ijr, PierreAuger:2023htc}, while AGNs and blazars also remain prime candidates for the diffuse background across multi-messenger channels \citep{Das:2020nvx, Murase:2011cy, Das:2025tfq}. While earlier studies by the Pierre Auger Collaboration focused on intermediate-scale searches using the 2FHL catalog \citep{PierreAuger:2018qvk}, our work investigates the large-scale dipole reconstruction using the Fermi-LAT 4LAC DR3 catalog alongside key nearby SBGs within $5$~Mpc (M82, NGC 253, NGC 4945, and M83) with their distance, flux, and attenuation modelling.

Crucially, our analysis demonstrates that neither the AGN population nor the SBG population alone can simultaneously reproduce both the observed Auger dipole amplitude and direction. To reconcile this discrepancy, we implement a hybrid framework that combines AGNs, nearby SBGs, and an isotropic background component. Across the investigated energy intervals, this hybrid model reveals that nearby SBGs provide the dominant anisotropic contribution.The remainder of 
this paper is organized as follows. In Section~\ref{sec:catalog}, we introduce 
the 4LAC AGN catalog and examine its large-scale source distribution, focusing 
on the gamma-ray energy fluxes and redshift properties. In Section 
\ref{sec:propagation}, we describe our composition-dependent propagation 
framework, which incorporates energy-dependent attenuation lengths and 
rigidity-dependent magnetic smearing. In section~\ref{results}, we discuss the results from
individual AGNs and SBG populations and culminating in our 
hybrid AGN+SBG anisotropy analysis across multiple energy intervals. Finally, 
we summarise our main findings and provide our concluding remarks in 
Section~\ref{sec:conclusion}.

\section{The 4LAC AGN Catalog and Source Distribution}\label{sec:catalog}

We use the \textit{Fermi}-LAT Fourth AGN Catalog Data, Release 3 
(4LAC DR3) \citep{Fermi-LAT:2022oww}, which provides a large sample of identified gamma-ray AGNs with 
measured sky positions, redshifts, and energy fluxes. From the catalog, we use 
the right ascension ($\alpha$), declination ($\delta$), redshift ($z$), and $F_{\gamma}$ denotes
the integral gamma-ray energy flux in the range from $100$ MeV to $100$ GeV. 
This integral gamma-ray energy flux is used as a proxy for source activity 
and is assumed to be 
related to the UHECR acceleration capability of the source. Gamma-ray bright 
AGNs are therefore expected to contribute more strongly to the observed UHECR 
flux. Following the removal of sources with missing or non-physical energy flux values, a total of 3407 valid AGNs are retained to form the parent sample for examining the large-scale source distribution prior to redshift selection and GZK attenuation. Subsequently, after filtering out sources with missing, zero, or non-physical redshifts, a refined sample of 1806 AGNs with valid measurements is finalized for the propagation and dipole analysis.

\begin{figure*}
\centering
\includegraphics[width=0.95\textwidth]{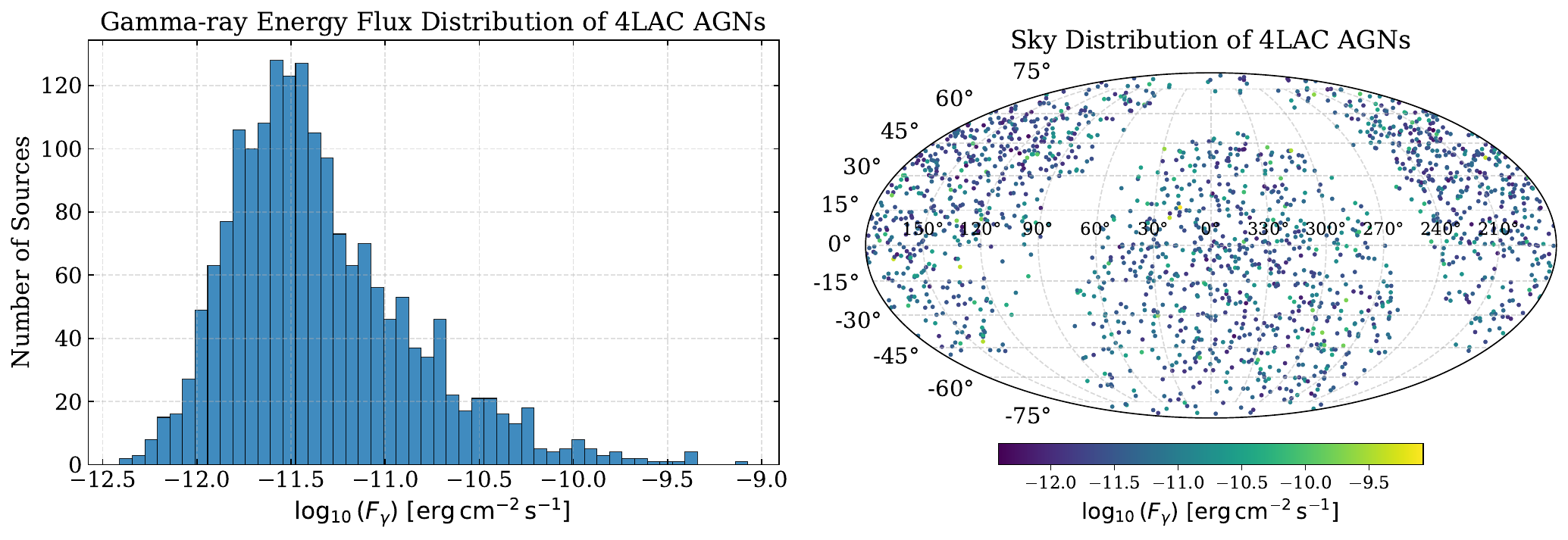}
\caption{Left: Distribution of the gamma-ray energy fluxes of 4LAC AGNs in the 
$100$ MeV--$100$ GeV energy range, showing that most sources are relatively 
faint while only a few are highly luminous. Right: Equatorial sky distribution
of the 4LAC AGNs in a Mollweide projection, color-coded by gamma-ray energy 
flux. The reduced source density near the Galactic plane is due to strong 
Galactic foreground emission, which limits the detection of extragalactic 
AGNs. The catalog serves as the source population for the UHECR dipole 
analysis.}
\label{fig:4lac_distribution}
\end{figure*}

Before performing the dipole anisotropy analysis, it is important to examine
the general properties of the gamma-ray AGN population used in this work. For
this purpose, Figure~\ref{fig:4lac_distribution} shows the basic distribution 
of the 4LAC AGN sample. The left panel presents the histogram of gamma-ray 
energy fluxes 
plotted in logarithmic scale. Most AGNs are relatively faint, while only a 
small number are very bright. This indicates that the source population is 
highly non-uniform, and the anisotropy is expected to be dominated mainly 
by a few nearby bright sources rather than by all sources equally. The right 
panel shows the equatorial sky distribution of these AGNs in a Mollweide 
projection using right ascension and declination coordinates. Each source is 
color-coded according to its gamma-ray energy flux. The apparent underdensity 
in some sky regions is caused by source incompleteness near the Galactic 
plane, where strong diffuse gamma-ray foreground emission from the Milky Way 
makes the identification of extragalactic AGNs difficult. This is a known 
observational selection effect and does not indicate a real absence of 
sources \citep{Fermi-LAT:2022oww}. Since UHECRs cannot propagate over arbitrarily large cosmological 
distances because of energy losses through interactions with the CMB, the 
redshift distribution of AGNs also becomes very important. In particular, the 
GZK effect strongly suppresses the contribution 
of distant sources and makes nearby extragalactic objects much more relevant 
for the observed dipole.

\begin{figure}
\centering
\includegraphics[width=0.46\textwidth]{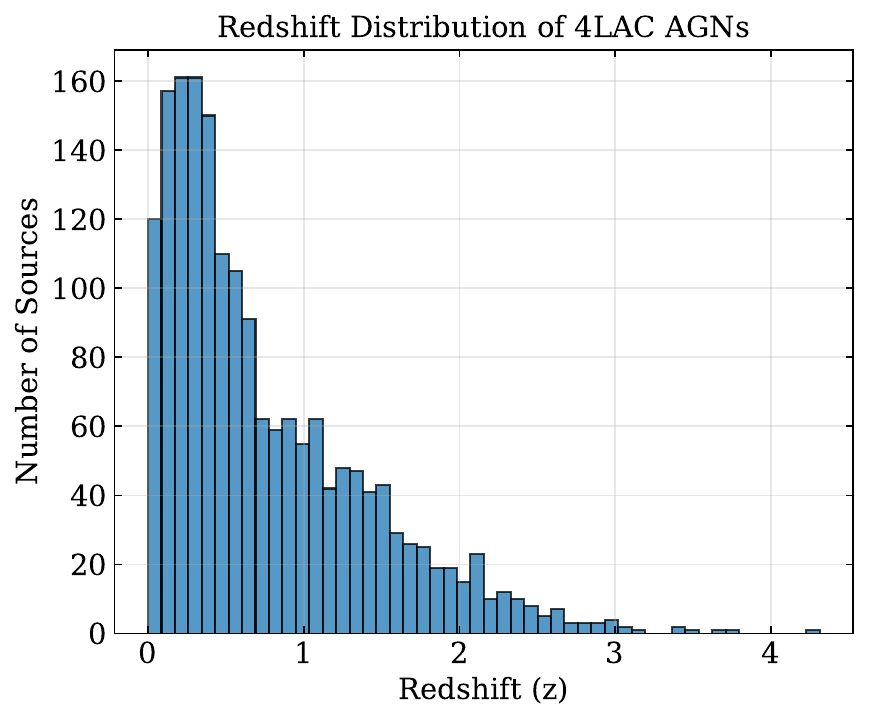}
\caption{Redshift distribution of 4LAC AGNs after removing sources with 
missing or non-physical redshifts. A total of $1806$ sources are retained for 
the analysis.}
\label{fig:redshift_distribution}
\end{figure}

We examine the redshift distribution of the same 4LAC sample. Figure~\ref{fig:redshift_distribution} shows 
the redshift distribution of these $1806$ AGNs. Most AGNs are located at 
moderate redshifts ($z\sim0.2$--$1$), while only a relatively small fraction 
lie in the nearby Universe. Since UHECRs experience energy-dependent 
attenuation during propagation, nearby sources generally contribute more 
strongly to the observed anisotropy than distant sources. To reconstruct the 
expected UHECR dipole each AGN is treated individually using its measured gamma-ray flux, 
redshift, and sky position, together with the energy and composition-dependent 
attenuation model. In the following section, we briefly discuss the propagation effect is discussed. 

\section{Propagation effect}
\label{sec:propagation}

To investigate the effect of UHECRs' composition on the observed dipole 
anisotropy, we perform a composition-dependent propagation analysis using the 
4LAC AGN catalog, as discussed above. The analysis incorporates 
energy-dependent attenuation during propagation together with 
rigidity-dependent magnetic deflections.


As mentioned already, the 4LAC DR3 catalog is used as the parent source 
population. For each source, the luminosity distance $d_i$ is calculated using the 
publicly available \textit{Planck18} cosmology code. The intrinsic dipole anisotropy is defined as \citep{Auger2018}
\begin{equation}\label{1}
d (E) = 3\,|\vec{D}(E)|,
\end{equation}
where 
\begin{equation}
\vec D(E) = \sum_{s=p, Z} f_s \vec D_{s} (E),
\end{equation}
where the composition fractions satisfy $\sum_s f_s = 1$. In the present analysis, we adopt the energy-dependent Auger composition fractions derived using the EPOS-LHC hadronic interaction model \citep{PierreAuger:2023xfc}.
Here, p refers protons while Z is for nuclei (A>1). The term $\vec D_{s} (E)$ is the weighted dipole vector that is constructed from the source distribution on the celestial sphere as \citep{Auger2018}
\begin{equation}
\vec{D}_{s}(E, Z) = \sum_i w_i (s) \hat{n}_i \exp \left[ - \frac{\sigma_i^2(E,s)}{2} \right].
\end{equation}
Here $\sigma(E, s)$ represents the effective smearing angle governed by the rigidity-dependent relation:
\begin{equation}
\sigma(E, s) = \sigma_0 s \left(\frac{E_0}{E} \right),
\end{equation}
where magnetic deflections during propagation in the Galactic magnetic field (GMF) complicate the original arrival directions of UHECRs \citep{PierreAuger:2018zqu}. While the characteristic displacement angle is known to be latitude-dependent, scaling approximately as $(\sin^2 b + 0.15)^{-1}$ \citep{diMatteo:2017dtg}, applying a purely symmetric, latitude-dependent dampening penalty to sources near the Galactic plane can introduce artificial geometric biasing into the calculated dipole vector. To avoid this and treat the deflections analytically, it is standard practice to conservatively apply an effective, sky-averaged uniform smearing model. Although the maximum expected displacement near the Galactic plane yields $\approx 4.7^\circ$ for $40$ EeV protons, we parameterize our effective uniform smearing by setting the baseline proton deflection scale to $\sigma_0 = 5^\circ$ at a reference energy of $E_0 = 60$ EeV \citep{Tanidis:2022jox}. 

The term $n_i$ is the unit vector toward each source with components,
\begin{equation}
\hat{n}_i =(
\cos\delta_i\cos\alpha_i,\,
\cos\delta_i\sin\alpha_i,\,
\sin\delta_i),
\end{equation}
here $\delta_i$ and $\alpha_i$ are right ascension
and declination respectively, for the $i^{\rm th}$ sources in the catalog as mentioned earlier. $w_i$ is the weight to each source and assigned according to
\begin{equation}
\begin{split}
    w_i(E, s) = F_{\gamma,i} \exp \left( - \frac{d_i}{\lambda_s(E)} \right) \\[10pt] \text{where } \lambda_s(E) = \begin{cases} \lambda_p(E), & \text{for proton} \\ \lambda_A(E), & \text{for nuclei} \end{cases}
\end{split}
\end{equation}
followed by normalization $\sum_i w_i = 1$. Here $\lambda$ denotes the energy-loss attenuation length and for protons, the total 
attenuation length is written as \citep{Aloisio:2010he}
\begin{equation}
\lambda_p^{-1}(E) = \lambda_{\gamma\pi}^{-1}(E) + \lambda_{ee}^{-1}(E) + \lambda_z^{-1},
\end{equation}
where the three terms correspond to the attenuation lengths for photopion 
production, pair production, and cosmological redshift losses, respectively. 
The redshift attenuation length is
\begin{equation}
\lambda_z = \frac{c}{H_0}.
\end{equation}
Following the analytic fits to proton energy-loss calculations in 
Ref.~\citep{harari} as
\begin{equation}
F(A,B,C,E) = A\exp(BE^C),
\end{equation}
where $A$, $B$ and $C$ are parameters for fitting with $E$ is the energy in EeV. The proton pair-production attenuation length is approximated 
as \citep{harari}
\begin{equation}
\lambda_{ee,p}(E)
= F(300,4.42,-0.6,E) + F(51,1.61,0.14,E),
\end{equation}
while the photopion attenuation length is given by \citep{harari}
\begin{equation}
\lambda_{\gamma\pi,p}(E) = F(11.5,686,-1.2,E).
\end{equation}

For nuclei heavier than protons, photodisintegration becomes the dominant 
energy-loss mechanism. The photodisintegration rate for a nucleus of mass 
number $A$ is written as \citep{Aloisio:2010he}
\begin{equation}
R_{A,i}
=
\frac{1}{2\Gamma^2} \int_0^\infty \frac{d\epsilon}{\epsilon^2} \frac{dn}{d\epsilon} \int_0^{2\gamma\epsilon} d\epsilon' \, \epsilon' \rho{A,i}(\epsilon'),
\end{equation}
where $\Gamma = {E}/{A m_p}$ is the Lorentz factor of the nucleus. 
The effective photodisintegration attenuation length is then written 
as \citep{Aloisio:2010he, harar_2015_prd}
\begin{equation}
\lambda_{\gamma d}^{-1} = \frac{1}{A} R_{A,\rm eff}.
\end{equation}
The pair-production attenuation length for nuclei is approximately scaled as 
\citep{Aloisio:2010he, harar_2015_prd}
\begin{equation}
\lambda_{ee,A}(E) = \lambda_{ee,p}(E) \left( \frac{A}{Z^2} \right),
\end{equation}
where $Z$ is the nuclear charge.
The total attenuation length for nuclei becomes \citep{Sarmah:2025urk}
\begin{equation}
\lambda_A^{-1}(E) = \lambda_{\gamma d}^{-1}(E) + \lambda_{ee,A}^{-1}(E) + \lambda_z^{-1}.
\end{equation}

\begin{figure}[ht!]
    \centering
    \includegraphics[width=\linewidth]{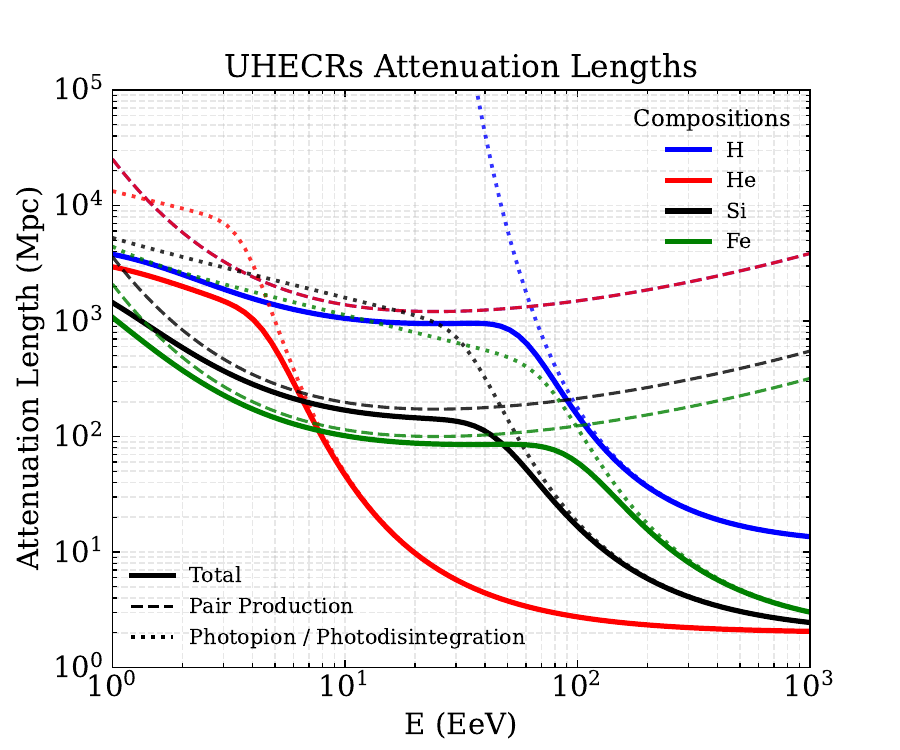}
   \caption{Energy-dependent attenuation lengths of UHECRs for different 
primary compositions: Proton (blue), Helium (red), Silicon (black), and Iron 
(green). Solid lines show the total attenuation length, dashed lines represent 
pair-production losses, and dotted lines indicate photopion production 
(protons) or photodisintegration (nuclei). The rapid decline of the dotted 
curves marks the onset of strong GZK energy losses.}
    \label{fig:attenuation_length}
\end{figure}
The resulting total attenuation lengths, alongside the isolated contributions 
from pair production, photopion production and photodisintegration losses, are 
illustrated in Figure \ref{fig:attenuation_length} as a function of primary 
energy. 

Finally, the dipole direction is reconstructed as \citep{Auger2018, PierreAuger:2014ati}
\begin{align} \label{16}
\delta_{\rm dip} &= \sin^{-1} \left(\frac{D_z}{d} \right), \qquad \alpha_{\rm dip} = \tan^{-1} \left(\frac{D_y}{D_x} \right).
\end{align}
In the following section, we discuss the results such as dipole amplitude, direction from AGN and SBGs.

\section{Single-Source Populations and the Hybrid Framework}\label{results}
In this section, we evaluate the large-scale dipole anisotropy predicted by individual candidate populations, specifically the Fermi-LAT 4LAC AGN and nearby SBGs as well as a hybrid framework incorporating their respective fluxes, distances, and propagation attenuation.

\subsection{Dipole Predictions from the 4LAC AGN Population}
Using the methodoogy shown in Eqns~\ref{1}-\ref{16}, the mixed-composition analysis for AGN-only source population is performed across diffrent energy intervals. In
$8-16$ EeV (the most statistically significant energy bin observed by the Auger) gives a dipole amplitude of $51.87\%$. The reconstructed 
dipole direction is found to be $(\alpha,\delta)=(167.3^\circ,73.6^\circ)$. In Table~\ref{tab:individual_sources}, other energy bins are also shown. The predicted dipole amplitude and dipole direction show a substantial disagreement with the Auger result. This suggests that the 4LAC AGN population alone is insufficient to fully explain the observed UHECR anisotropy pattern.

\subsection{Dipole Predictions from the nearby SBGs} \label{sec:starburst}
For the SBG analysis, we restrict our source selection to SBGs within $5$ Mpc, for which UHECR propagation attenuation is expected to be relatively small \citep{apj935}. Within this regime, we consider the four dominant SBGs frequently associated with UHECR anisotropy studies: M82, NGC~253, NGC~4945, and M83 as shown in Table~\ref{tab:sbg_properties}. As demonstrated by the Auger collaboration \citep{PierreAuger:2018qvk}, when local SBGs are weighted by their radio continuum emission and corrected for UHECR propagation attenuation, these four sources stand out as the primary theoretical contributors \citep{apj935}. Under standard composition and attenuation models, M82 ($\sim 24\%$), NGC~253 ($\sim 21\%$), NGC~4945 ($\sim 19\%$), and M83 ($\sim 8\%$) collectively account for the vast majority ($>70\%$) of the total expected UHECR flux at $1-80$ EeV. Furthermore, when folding in the directional exposure of the Southern Hemisphere-based Auger Observatory, the expected anisotropy signal is overwhelmingly driven by NGC~4945 ($\sim 39\%$), NGC~253 ($\sim 36\%$), and M83 ($\sim 13\%$) while M82's observable contribution to anisotropy drops to $\sim 0.2\%$ strictly due to its high northern declination \citep{apj935}. The vast majority of the other remaining SBGs in the catalog \citep{apj935} yield an expected null contribution of $\sim 0\%$ due to a combination of distance, lower intrinsic flux, and limited detector exposure.
\begin{table}[htbp]
\centering
\caption{Coordinates, distances, and fluxes of the key nearby starburst galaxies \citep{apj935}.}
\label{tab:sbg_properties}
\begin{tabular}{lcccc}
\hline
Source & $\alpha$ ($^\circ$) & $\delta$ ($^\circ$) & $d$ (Mpc) & Flux weight (\%) \\
\hline
M82       & 148.97 & $+69.68$ & 3.61 & 18.6 \\
NGC 253   &  11.89 & $-25.29$ & 3.70 & 13.6 \\
NGC 4945  & 196.37 & $-49.47$ & 3.47 & 16 \\
M83       & 204.25 & $-29.87$ & 4.90 & 6.3 \\
\hline
\end{tabular}
\end{table}

Using these SBGs, we perform a mixed-composition dipole analysis using Eqns~\ref{1}-\ref{16}. The resulting mixed-composition dipole amplitude 
(for $8-16$ EeV) is found to be  $30.49\%$, which is close to the large-scale dipole amplitude reported by the Auger. The reconstructed dipole direction is obtained as $(\alpha,\delta)=(48.0^\circ,-65.9^\circ)$. Table~\ref{tab:individual_sources} lists the amplitude, 
RA and DEC, separation for other energy intervals. We see that SBGs also not solely reproduce the amplitude and direction of Auger.
\begin{table*}
\small
\caption{Energy-dependent dipole properties obtained for the individual AGN and SBG source scenarios.}
\label{tab:individual_sources}
\hspace*{-1.6cm}
\begin{tabular}{c|cc|ccc|ccc}
\hline
\multirow{2}{*}{Energy Bin} & \multicolumn{2}{c|}{Auger} & \multicolumn{3}{c|}{AGN Scenario} & \multicolumn{3}{c}{SBG Scenario} \\
\cline{2-9}
& Amp (\%) & (RA, Dec) ($^\circ$) & Amp (\%) & RA ($^\circ$) & Dec ($^\circ$) & Amp (\%) & RA ($^\circ$) & Dec ($^\circ$) \\
\hline
4--8~EeV   & $0.6^{+0.7}_{-0.3}$ & $(80 \pm 60, -75^{+17}_{-8})$   & 38.68 & 172.0 & $+72.2$ & 23.98 & 48.0 & $-65.9$ \\
8--16~EeV  & $5.8^{+1.3}_{-1.1}$ & $(104 \pm 11, -24^{+12}_{-13})$ & 51.87 & 167.3 & $+73.6$ & 30.49 & 48.0 & $-65.9$ \\
16--32~EeV & $6.5^{+2.5}_{-1.8}$ & $(82 \pm 20, -50^{+15}_{-14})$  & 58.49 & 144.9 & $+69.6$ & 28.42 & 47.7 & $-65.7$ \\
$>32$~EeV  & $8^{+5}_{-3}$       & $(115 \pm 35, -46^{+28}_{-26})$ & 38.73 & 166.2 & $+13.4$ & 28.82 & 46.5 & $-64.9$ \\
\hline
\end{tabular}
\end{table*}
Nevertheless, the dominance of nearby southern SBGs naturally shifts the 
dipole toward negative declinations, indicating that local SBGs 
may provide an important contribution to the observed large-scale anisotropy. These results therefore motivate a hybrid scenario in which both AGNs and nearby SBGs contribute to the observed UHECR sky distribution.

\subsection{Hybrid Anisotropy Framework}

Since neither source class acting alone can simultaneously reproduce both the dipole amplitude and celestial direction measured by the Auger across the observed energy range, this motivates a hybrid scenario in which distant AGNs and nearby SBGs jointly contribute to the observed UHECR sky distribution. To investigate this possibility, we construct a hybrid dipole 
model by combining the AGN and SBG dipole vectors together with an isotropic 
background component inspired from \citep{apj935, Harari:2015mal, mollerach2022} as $\vec{d}_{\text{total}} = f_{\text{AGN}}\,\vec{d}_{\text{AGN}} + f_{\text{SBG}}\,\vec{d}_{\text{SBG}} + f_{\text{iso}}\,\vec{d}_{\text{iso}}$ \footnote{In Ref.~\citep{apj935, Harari:2015mal, mollerach2022}, authors considered an isotropic component ($i$ or $\alpha$), it is not the same component as we consider here as $f_{\text{AGN}}$ or $f_{\text{SBG}}$.}
where $\vec{d}_{\text{AGN}}(E) = 3\,\vec{D}_{\text{AGN}}(E)$ and $\vec{d}_{\text{SBG}}(E) = 3\,\vec{D}_{\text{SBG}}(E)$ are the individual dipole vectors defined with the standard normalisation factor. The isotropic fraction $f_{\text{iso}}$ is essential as
in nature, a significant portion of the observed CRs comes from unresolved, faint, and distant cosmological sources, or from primary nuclei whose arrival directions have been completely randomised by intervening turbulent magnetic fields. This component forms a diffuse, isotropic or monopole background. Due to spherical symmetry, its dipole vector vanishes identically ($\vec{d}_{\text{iso}} = \vec{0}$), so the total dipole is given by
\begin{equation}
\vec{d}_{\text{total}} = f_{\text{AGN}}\,\vec{d}_{\text{AGN}} + f_{\text{SBG}}\,\vec{d}_{\text{SBG}}
\end{equation}
Because the individual AGN and SBG dipole vectors point toward distinctly different directions on the celestial sphere, the total hybrid dipole amplitude $d_{\text{total}}$ is obtained via 3D vector addition as
\begin{equation}
\begin{aligned}
d_{\text{total}}
&= \left\lVert \vec{d}_{\text{total}} \right\rVert \\
&= \sqrt{\vphantom{\frac{1}{1}}
\begin{aligned}
&(f_{\text{AGN}}d_{\text{AGN}})^2
+(f_{\text{SBG}}d_{\text{SBG}})^2\\[-2pt]
&\quad +2f_{\text{AGN}}f_{\text{SBG}}
d_{\text{AGN}}d_{\text{SBG}}\cos\theta_{\text{AS}}
\end{aligned}}
\end{aligned}
\end{equation}
where $d_{\text{AGN}} = \Vert{}\vec{d}_{\text{AGN}}\Vert{}$ and $d_{\text{SBG}} = \Vert{}\vec{d}_{\text{SBG}}\Vert{}$ are the single-population dipole amplitudes listed in Table~\ref{tab:individual_sources}, and $\theta_{\text{AS}}$ is the angular separation between the individual AGN and SBG dipole directions on the sky.
The fraction $f_{\rm iso}$ is calculated as $f_{\rm iso} = 1 - f_{\rm AGN} - f_{\rm SBG}$

The analysis is performed simultaneously for the $4$--$8$ EeV, $8$--$16$ EeV, 
$16$--$32$ EeV, and $>32$ EeV energy intervals using the corresponding 
Auger dipole amplitudes and directions as observational targets. For each 
energy bin, the attenuation-suppressed AGN and SBG dipole vectors are first 
calculated using the mixed-composition framework described previously. 
To determine the optimal source fractions, we perform a parameter scan over 
$f_{\rm AGN}$ and $f_{\rm SBG}$ in the range $0\leq f\leq1$ with a step size 
of $0.01$.
To evaluate the model against observational measurements from the Auger, the best-fit configuration $(f_{\text{AGN}}, f_{\text{SBG}})$ is determined by minimizing a joint statistical $\chi^2$ metric that simultaneously accounts for amplitude discrepancies and directional deviations,
\begin{equation}
\chi^2 = \left( \frac{d_{\text{model}} - d_{\text{Auger}}}{\sigma_{d, \text{Auger}}} \right)^2 + \left( \frac{\theta_{\text{sep}}}{\sigma_{\theta, \text{Auger}}} \right)^2,
\end{equation}
where $d_{\text{model}}$ and $d_{\text{Auger}}$ denote the predicted and observed dipole amplitudes, respectively, and $\sigma_{d, \text{Auger}}$ is the statistical uncertainty on the measured dipole amplitude. The second term implies directional misalignment, where $\theta_{\text{sep}}$ is the angular separation on the unit sphere between the model and observed dipole directions, and $\sigma_{\theta, \text{Auger}}$ represents the positional uncertainty of the observed dipole orientation \citep{Auger2018}. In the global fitting analysis, we obtained the total chi-square as 2.06.

The angular separation $\theta_{\text{sep}}$ is calculated using the spherical dot product:
\begin{align}
\cos\theta_{\text{sep}} = &\sin\delta_{\text{model}}\sin\delta_{\text{Auger}} +\\ \nonumber 
&\cos\delta_{\text{model}}\cos\delta_{\text{Auger}}\cos(\alpha_{\text{model}} - \alpha_{\text{Auger}}),
\end{align}
\begin{equation}
\theta_{\text{sep}} = \arccos\left(\cos\theta_{\text{sep}}\right) \times \left(\frac{180^\circ}{\pi}\right),
\end{equation}
The uncertainties $\sigma_{d,\mathrm{Auger}}$ and
$\sigma_{\theta,\mathrm{Auger}}$ are adopted from the corresponding
Auger measurements for each energy interval. Both
$\theta_{\mathrm{sep}}$ and $\sigma_{\theta,\mathrm{Auger}}$ are
expressed in degrees. The joint metric therefore favors source
fractions that simultaneously reproduce the observed dipole amplitude
and minimize the angular separation from the measured dipole direction. The resulting best-fit parameters 
for the different energy intervals are summarized in Table~\ref{tab:hybrid_fit}.

\begin{table*}[ht]
\centering
\caption{Best-fit hybrid AGN+SBG anisotropy parameters for different energy intervals.}
\label{tab:hybrid_fit}
\hspace*{-2.4cm}
\begin{tabular}{cccccccc}
\hline
Energy Bin & $f_{\rm AGN}$ & $f_{\rm SBG}$ & $f_{\rm iso}$ & Amplitude & RA & Dec & Separation \\
 & (\%) & (\%) & (\%) & (\%) & ($^\circ$) & ($^\circ$) & ($^\circ$) \\
\hline

$4$--$8$ EeV & 0 & 3 & 97 & 0.72 & 76.2 & $-66.1$ & 8.98\\

$8$--$16$ EeV
& 23
& 45
& 33
& 5.03
& 104.1
& $-25.1$
& 7.10
\\

$16$--$32$ EeV
& 15
& 47
& 39
& 7.05
& 92.7
& $-43.5$
& 9.78
\\

$>32$ EeV
& 20
& 41
& 39
& 11.08
& 117.8
& $-60.20$
& 14.30
\\

\hline
\end{tabular}
\end{table*}

The hybrid model successfully reproduces both the amplitude and directional 
evolution of the observed large-scale dipole anisotropy. In particular, the 
agreement improves significantly at higher energies, where the angular 
separation between the reconstructed dipole and the observed Auger dipole 
decreases to only a few degrees.

The results indicate that nearby SBGs provide the dominant 
contribution to the observed anisotropy at high energies, while AGNs 
contribute subdominantly together with an approximately isotropic background 
component. Furthermore, the reconstructed dipole progressively shifts toward 
southern declinations with increasing energy, consistent with the growing 
influence of nearby southern SBGs such as NGC~253 and NGC~4945. This trend is 
qualitatively consistent with the energy-dependent dipole evolution reported 
by the Auger.

\begin{figure*}
\centering
\includegraphics[scale=0.4]{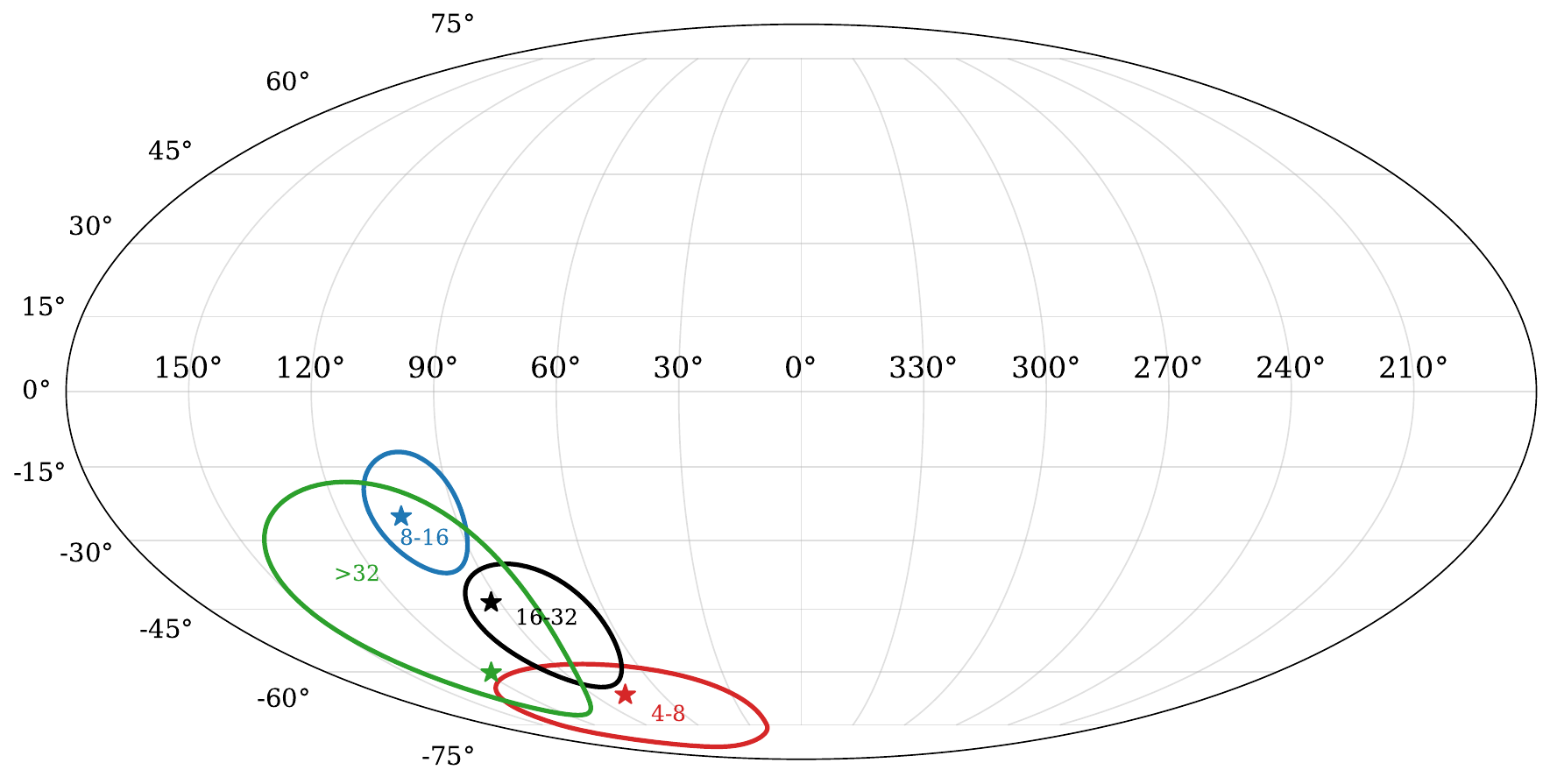}
\caption{Energy evolution of the reconstructed hybrid AGN+SBG dipole direction 
in equatorial coordinates for the $4$--$8$ EeV, $8$--$16$ EeV, $16$--$32$ EeV, 
and $>32$ EeV energy intervals. The coloured contours represent the 
corresponding Auger dipole-direction uncertainty regions, while the star 
symbols denote the reconstructed dipole directions obtained from the best-fit 
hybrid source fractions listed in Table~\ref{tab:hybrid_fit}. The 
reconstructed dipole progressively shifts toward southern declinations with 
increasing energy, reflecting the growing dominance of nearby southern 
starburst galaxies at the highest energies. The agreement between the 
reconstructed and observed dipole directions improves significantly with 
energy.}
\label{fig:hybrid_map}
\end{figure*}
\begin{figure}[ht]
\centering
\includegraphics[scale=0.5]{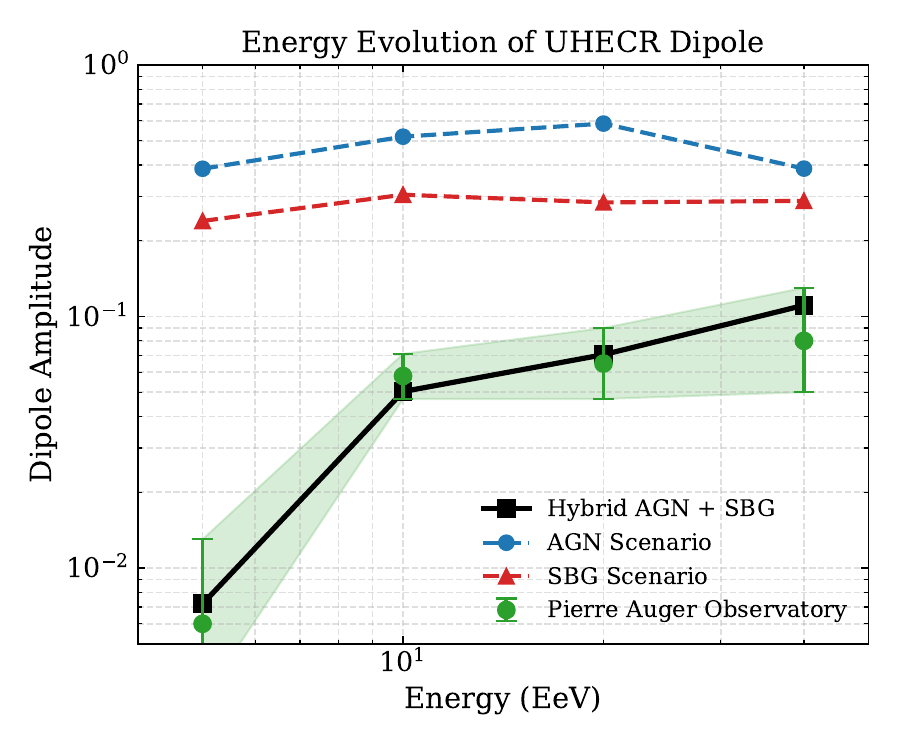}
\caption{Energy evolution of the UHECR dipole amplitude in the hybrid AGN+SBG 
framework compared with the Auger measurements.}
\label{fig:dipole_evolution}
\end{figure}

The best-fit hybrid anisotropy parameters obtained for the different energy 
intervals are summarized in Table~\ref{tab:hybrid_fit}, while the 
corresponding dipole-direction evolution on the sky is shown in 
Fig.~\ref{fig:hybrid_map}. The reconstructed dipole directions are found to 
evolve systematically toward southern declinations with increasing energy, in 
qualitative agreement with Auger observations.

Looking at the best-fit parameters across the different energy bins, we can observe how the source fractions change with energy.
At the lowest energy interval ($4$--$8$ EeV), the model is mostly made up of the isotropic component ($97\%$). The SBG contribution is very small ($7\%$), and there is no AGN contribution ($0\%$). This creates a small reconstructed amplitude of $0.72\%$ and a angular separation of $8.98^\circ$ from the observed Auger dipole. A high isotropic fraction is expected at these lower energies because magnetic deflections are stronger and particles can travel from further away, which blurs the arrival directions.In the $8$--$16$ EeV interval, the source fractions change noticeably. The SBG contribution rises to $45\%$, and the AGN contribution increases to $23\%$. The reconstructed dipole amplitude is $5.03\%$, which is close to the Auger value of approximately $5.8^{+1.3}_{-1.1}\%$. The reconstructed dipole direction is located at $(\alpha,\delta)=(104.1^\circ,-25.1^\circ)$, giving an angular separation of $7.10^\circ$ from the observed Auger dipole direction. We adopt the observed Auger dipole amplitude and direction directly from Ref.~\citep{Auger2018}. For the $16$--$32$ EeV interval, the source fractions show a small shift. The AGN contribution decreases to $15\%$, while the SBG fraction continues to increase slightly to $47\%$. The reconstructed dipole amplitude is $7.05\%$, which is close to the observed Auger amplitude in this energy range. The corresponding dipole direction is obtained at $(\alpha,\delta)=(92.7^\circ,-43.5^\circ)$, resulting in an angular separation of $9.78^\circ$ from the Auger dipole direction. At the highest energies ($>32$ EeV), the AGN contribution increases slightly to $20\%$, the SBG contribution decreases slightly to $41\%$. The reconstructed dipole amplitude reaches $11.08\%$, which is near the observed Auger value of about $8^{+5}_{-3}\%$. The reconstructed dipole direction moves to $(\alpha,\delta)=(117.8^\circ,-60.20^\circ)$, with an angular separation of $14.30^\circ$ from the observed Auger dipole.
The coloured contours in Fig.~\ref{fig:hybrid_map} represent the corresponding 
Auger dipole-direction uncertainty regions, while the star symbols denote the 
reconstructed dipole directions obtained from the best-fit hybrid source 
fractions listed in Table~\ref{tab:hybrid_fit}. The directional evolution 
shown in the figure indicates that the dipole progressively migrates toward 
the southern sky with increasing energy. This behaviour is naturally explained 
by the increasing dominance of nearby southern SBGs, particularly NGC~253 and 
NGC~4945, whose contributions become increasingly important at high energies 
due to the strong attenuation of distant extragalactic sources. Overall, the 
hybrid AGN+SBG framework successfully reproduces both the amplitude and 
directional evolution of the observed UHECR dipole anisotropy across the 
considered energy intervals. We note that the quoted AGN, SBG, and isotropic contributions represent our best-fit values; however, variations of approximately 
$\pm 1–2$ in these contributions may also provide comparable fits to the Auger anisotropy data across the individual energy intervals.

Figure~\ref{fig:dipole_evolution} shows the energy evolution of the 
reconstructed dipole amplitude for the hybrid AGN+SBG framework together with 
the corresponding Auger measurements. The hybrid model successfully reproduces 
the observed increase in the dipole amplitude with energy over the considered 
energy intervals.
The dashed curves represent the individual AGN-only and SBG-only contributions 
after including attenuation effects, mixed composition, and rigidity-dependent 
magnetic smearing. Although both source classes generate sizeable anisotropies,
neither of them alone is able to reproduce the full energy evolution observed 
by the Auger. In particular, the AGN-only contribution produces a 
comparatively weaker anisotropy at high energies, while the SBG-only 
contribution predicts a stronger dipole evolution dominated by nearby sources.
The solid curve corresponds to the best-fit hybrid configuration obtained from 
the combined AGN and SBG contributions together with an isotropic background 
component. The inclusion of both source populations significantly improves the 
agreement with the Auger data across the entire energy range. At low energies, 
the anisotropy remains strongly suppressed due to the dominant isotropic 
background contribution, whereas at higher energies the nearby SBGs 
progressively dominate the dipole signal, leading to a rapid increase in the 
anisotropy amplitude.
The points with asymmetric error bars denote the Auger Rayleigh dipole 
measurements from the SD1500 data set, while the shaded regions 
indicate the corresponding observational uncertainties. Overall, the hybrid 
scenario provides a consistent description of both the amplitude and 
directional evolution of the observed large-scale UHECR anisotropy.

\section{Summary and Conclusion}
\label{sec:conclusion}

In this work, we have investigated the origin of the large-scale dipole 
anisotropy of UHECRs observed by the Auger at energies 
above $8$ EeV. By incorporating a comprehensive framework that includes 
energy-dependent propagation losses (such as the GZK effect and 
photodisintegration), mixed-composition assumptions based on Auger data, and 
rigidity-dependent magnetic deflections, we evaluated the individual and 
combined contributions of gamma-ray AGNs from the Fermi-LAT 4LAC DR3 catalog 
and nearby SBGs.

Our independent analyses of single-source populations revealed that neither 
AGNs nor SBGs alone can simultaneously reproduce the observed dipole amplitude 
and direction. Specifically, the AGN-only model in the $8-16$ EeV range 
yielded a highly overestimated dipole amplitude of $51.87$\%. Conversely, the 
SBG-only scenario yielded an amplitude of $30.49$\%. These results demonstrate that a single source class is insufficient to explain the full anisotropy pattern.

To resolve these discrepancies, we constructed a hybrid model combining the 
AGN and SBG dipole vectors alongside an isotropic background component. This 
hybrid scenario successfully reconciles our theoretical predictions with the 
observational data across multiple energy bins.
The hybrid model successfully reproduces the observed Auger dipole across all energy bins, with SBGs providing the dominant anisotropic component. In the statistically prominent 8–16 EeV bin, the model yields a dipole amplitude of 5.03\% and an angular separation of only $7.10^\circ$ from the Auger direction, requiring a 45\% SBG and 23\% AGN contribution. At lower energies, strong magnetic deflections naturally drive a large isotropic fraction ($\approx 97\%$), while at higher energies the reconstructed amplitudes and directions remain consistently aligned with Auger measurements.
Recently Auger has reported a smaller fraction $\alpha \approx 9\% - 19\%$ that searches measures the  highly localized flux excess clustered tightly around galaxy positions on intermediate scales ($\Theta \approx 15^\circ - 18^\circ$). Our analysis is different than the Auger analysis \citep{apj935}. We also note that our obtained percentages represent the best-fit values; small variations of about $\pm 1-2\%$ in the source contributions still provide a comparable fits to the Auger data.

\section*{Acknowledgement}
UDG is thankful to the Inter-University Centre for Astronomy and Astrophysics 
(IUCAA), Pune, India for the Visiting Associateship of the institute.

\bibliography{main}{}
\bibliographystyle{aasjournalv7}

\end{document}